\documentclass[twocolumn,aps,floatfix,superscriptaddress]{revtex4}
\pdfoutput=1 %This is for the ArXiv submission only
\usepackage{amsmath,amssymb,eucal,graphicx,float,epstopdf}

\begin{document}
\title{Kinetics of Deposition in the Diffusion-Controlled Limit}
\author{P.~L.~Krapivsky}
\affiliation{Department of Physics, Boston University, Boston, Massachusetts 02215, USA}

\begin{abstract} 
The adsorption of particles diffusing in a half-space bounded by the substrate and irreversibly sticking to the substrate upon contacts is investigated. We show that when absorbing particles are planar disks diffusing in the three-dimensional half-space, the coverage approaches its saturated ``jamming" value as $t^{-1}$ in the large time limit [generally as $t^{-1/(d-1)}$ when the substrate is $d$ dimensional and $d>1$, and as $e^{-t/\ln(t)}$ when $d=1$]. We also analyze the asymptotic behavior when particles are spherical and when particles are planar aligned squares. 
\end{abstract}

\maketitle

\section{Introduction}

The deposition of suspended particles onto substrates plays an important role in physics, chemistry and biology  \cite{Tor,Adam,Newby13}.  Adhesion of colloidal particles and proteins on clean substrates is a classic example. There are also numerous applications, e.g. in nanotechnology \cite{Elimelech02,Kuznar03,Floro06,Yaish13}. A comprehensive description of the deposition process requires the understanding of the evolution of a strongly interacting infinite-particle system. Indeed, suspended particles diffuse and directly interact with each other through exclusion. Further, the motion of suspended particles causes long-ranged hydrodynamic inter-particle interactions.  The process of attaching to the substrate is also very complicated---the particles may rotate, the shape of the particles plays an important role, etc. Even when the volume fraction occupied by suspended particles is small and the inter-particle interactions in the solution are ignored, little is known theoretically. It is therefore customary to maximally simplify the problem. Following this tradition we focus on dilute systems, ignore rotations and consider particles of a few simple shapes. Most studies also ignore diffusion. In contrast, our major goal is to probe the influence of diffusion of suspended particles. Specifically, we consider the diffusion-controlled limit in which adhesion occurs instantaneously and is assumed to be irreversible. 

A comprehensive treatment of the diffusion-controlled deposition problem is beyond the reach of analytical approaches, so we shall employ heuristic arguments. One particular situation amenable to heuristic treatment is when the particles are planar disks diffusing in a half-space above the flat substrate. Disks are assumed to remain parallel to the substrate and when a disk touches the substrate, it irreversibly adheres to it. The overlapping of disks on the substrate is forbidden. The substrate eventually reaches a jammed state that cannot accommodate additional disks. What fraction of the substrate is covered in the jammed state? What is the temporal evolution in the vicinity of the jammed coverage? Are there correlations in particle positions in the jammed state? 

Some progress in answering such questions has been achieved in the setting which ignores diffusion and mimics  the solution as a reservoir of particles. This framework is known as the random sequential adsorption (RSA). The RSA model postulates that the deposition events are random: If the new particle does not overlap with already deposited ones, it sticks to the substrate; otherwise, the deposition event is discarded. 

The RSA model was introduced a long time ago \cite{F39,R58,Pal60} and it is fairly well-understood (see \cite{B-Privman,Evans,TT,C-Privman,book,Itoh} for a review), although analytical solutions have been established only in the case of the one-dimensional substrate (when disks become segments). For the RSA of balls the basic properties of the jammed state have not been determined analytically, e.g. the jamming coverage $\rho_\text{jam}$ is unknown, see \cite{ZT13}  for accurate numerical results in dimensions $2\leq d\leq 8$. The asymptotic approach to the jamming coverage is known \cite{Feder80,P80,Sw81}, namely $\rho_\text{jam}-\rho(t)\sim t^{-1/2}$ for the RSA of disks. This result admits a generalization to arbitrary dimension \cite{P80,Sw81}: 
\begin{equation}
\label{alg:approach}
\rho_\text{jam}-\rho(t)\sim t^{-\sigma}
\end{equation}
for $t\gg 1$ with  jamming exponent $\sigma=1/d$ for the RSA of balls onto the $d$ dimensional substrate.

Random packings generated by RSA have been used in various fields ranging from soft matter \cite{TS10} to telecommunication \cite{Coffman94,Gnedin01}. The application of RSA to deposition processes is more questionable as the diffusive motion of particles in solution above the substrate is clearly important. Fortunately, the effect of this motion on the jamming coverage is rather small \cite{T:JCP92,Ramsden97}, but the kinetic approach to the jamming coverage is definitely very different as we shall demonstrate below. 

The outline of the paper is as follows. In Sec.~\ref{sec:ADD} we show that for the adsorption of diffusing disks the asymptotic approach to the jamming coverage is also algebraic, and we determine the jamming exponent. We then analyze the diffusion-controlled deposition process when the diffusing particles are aligned planar squares (Sec.~\ref{sec:ADSQ}) or spheres (Sec.~\ref{sec:ADS}). In Sec.~\ref{sec:DS} we briefly discuss how one can include short-ranged interactions, particularly exclusion volume interactions, between diffusing particles. We conclude with a discussion (Sec.~\ref{sec:Discussion}).

\section{Adsorption of Diffusing Disks}
\label{sec:ADD}

The basic assumptions underlying the deposition process of diffusing planar discs are as follows:
\begin{enumerate}
\item Disks freely diffuse (no interactions) in the solution and they remain parallel to the substrate. 
\item Adsorbed disks do not overlap, do not desorb from the substrate, and do not diffuse on the substrate.
\item The solution is dilute, so we ignore interactions between disks in the solution. 
\end{enumerate}

Even for planar particles one can consider the general situation when they do not remain parallel to the substrate. Furthermore, in the case of planar particles which remain parallel to the substrate, the analysis becomes more challenging when particles are different from disks. We shall comment on these more complicated systems later, e.g. we consider the diffusion-controlled adsorption of aligned planar squares in Sect.~\ref{sec:ADSQ}, but in this section we consider diffusing planar disks irreversibly sticking to the substrate. 

\begin{figure}[ht]
\begin{center}
\includegraphics[width=0.44\textwidth]{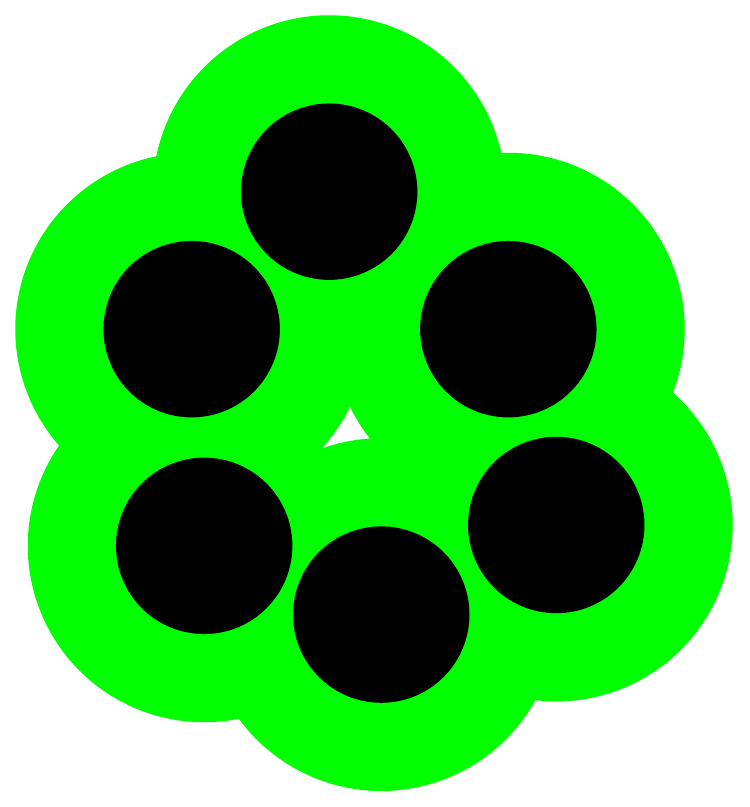}
\caption{Illustration of a target zone (white). Each adsorbed disk (black) is surrounded by an exclusion zone of twice larger radius (green). In the long-time limit target zones are triangular with arc-shaped sides whose radius is twice that of the disks. Target zones are far away from each other and hence they effectively do not interact in the long-time limit.}
\label{Fig:zone}
  \end{center}
\end{figure}

Let us start with the most important case of adsorption on the plane ($d=2$). In the long-time limit there is little room for new disks, namely their centers can land into disconnected target zones which are roughly triangular in shape (more precisely, they have arc-shaped sides whose radius is twice that of the disks; see Fig.~\ref{Fig:zone}). The separation between remaining target zones grows indefinitely and in the long-time regime one can ignore interactions between target zones. This greatly simplifies the determination of the asymptotic behavior of the coverage and it was the chief idea of Pomeau \cite{P80} and Swendsen \cite{Sw81} in the case of the RSA model that led to \eqref{alg:approach} with $\sigma=1/d$. The same holds in our case, although the elimination of the target zones proceeds in a different manner. 

Denote by $c(\ell,t)$ the density of target zones of linear size $\ell$. In the long time limit almost all target zones are very small, $\ell\ll R$ where $R$ is the radius of disks, well-separated and effectively non-interacting, so the density $c(\ell,t)$ decays according to the linear rate equation 
\begin{equation}
\label{eq:ct2}
\frac{dc(\ell,t)}{dt}\simeq -K_2 c(\ell,t)
\end{equation}
The amplitude $K_2$ depends on the diffusion coefficient $D$, the size $\ell$ of the target zone and the density $n_\infty$ of the disks far away from the target. On pure dimensional grounds one can write $K_2=n_\infty D \ell\, \mathcal{K}(n_\infty \ell^2)$, where $\mathcal{K}(n_\infty \ell^2)$ is some function of the (dimensionless) fraction of uncovered area $n_\infty \ell^2$. However, we consider dilute solutions, so the dependence of $K_2$ on $n_\infty$ must be linear. (Also $n_\infty \ell^2\to 0$ in the long time limit, so one can replace $\mathcal{K}(n_\infty \ell^2)\to \mathcal{K}(0)$ and obtain the same result.) Thus $K_2=A_2n_\infty D \ell$ where $A_2$ is a numerical factor that cannot be determined by dimensional analysis. Combining \eqref{eq:ct2} and $K_2=A_2n_\infty D \ell$ we deduce 
\begin{eqnarray}
\label{ct:2d}
\rho_\text{jam}-\rho(t) &\sim& \int_0^R \frac{d\ell}{R}\, c(\ell,t) \sim \int_0^R \frac{d\ell}{R}\, e^{-A_2 n_\infty D \ell t}\nonumber \\
                                    &\sim& R^{-1} (n_\infty D t)^{-1}
\end{eqnarray}

For the $d$ dimensional substrate with $d>1$, an analog of \eqref{eq:ct2} is
\begin{equation}
\label{eq:ctd}
\frac{dc}{dt}\simeq - K_d c, \quad K_d=A_d n_\infty D \ell^{d-1}
\end{equation}
where we used again dimensional analysis to fix $K_d$. Similarly to \eqref{ct:2d} 
we deduce 
\begin{equation}
\label{ct:above}
\rho_\text{jam}-\rho(t) \sim  R^{-1}(n_\infty Dt)^{-\frac{1}{d-1}}
\end{equation}
The jamming exponent $\sigma=1/(d-1)$ characterizing the diffusion-controlled deposition process is larger than the jamming exponent $\sigma=1/d$ characterizing the RSA.

The jamming exponent $\sigma=1/(d-1)$ predicted by \eqref{ct:above} becomes infinite if $d=1$. This means that the algebraic approach \eqref{alg:approach} is no longer valid. (The disks are segments when $d=1$, they remain parallel to the one-dimensional substrate, diffuse in the two-dimensional half-space and adhere to the substrate upon touching it.) To be able to treat the diffusion-controlled deposition onto the one-dimensional substrate and to establish the dependence $K_d \sim n_\infty D \ell^{d-1}$ for $d>1$ without dimensional analysis, we now describe a more comprehensive approach. We still want to establish the asymptotic behavior, so we can assume that target zones are small, $\ell\ll R$, and well-separated. For dilute suspensions we can also neglect the exclusion volume interactions between the disks. Thus we can treat the centers of the disks as non-interacting Brownian point particles. The triangular zone is eliminated when a point particle touches it. We can disregard the reflection boundary condition on the substrate by considering a ``two-sided" problem, namely a planar $d-$dimensional target zone at $z=0$ in $\mathbb{R}^{d+1}$. 

Consider a single not necessarily planar target $\mathcal{B}$. The density $n({\bf r}, t)$ of Brownian particles exterior to the target satisfies the diffusion equation
\begin{equation}
\label{nrt-reaction}
 \frac{\partial n}{\partial t} = D\nabla^2 n
\end{equation}
Far away from the target  
\begin{subequations}
\begin{equation}
\label{nrt-inf}
n({\bf r}\to\infty, t)=n_\infty 
\end{equation}
and absorption on the boundary $\partial \mathcal{B}$ of the target yields the boundary condition
\begin{equation}
\label{nrt-BC}
n({\bf r}\in \partial \mathcal{B}, t) =0
\end{equation}
\end{subequations}
In the long-time limit, the density approaches the steady state when $d>1$; see, e.g., \cite{book,OTB89}. The linearity of the governing equations \eqref{nrt-reaction}--\eqref{nrt-BC} implies that the density is proportional to $n_\infty$. It is convenient to write the stationary density in the form $n({\bf r}, t\to\infty)=n_\infty[1-\phi(\mathbf{r})]$. Plugging this form into the above governing equations we find that $\phi(\mathbf{r})$ obeys the Laplace equation
\begin{equation}
\label{Laplace}
\nabla^2 \phi = 0
\end{equation}
while the boundary conditions \eqref{nrt-inf}--\eqref{nrt-BC} become
\begin{equation}
\label{phi:BC}
\phi({\bf r}\in \partial \mathcal{B}) =1, \quad 
\phi({\bf r}\to\infty) = 0
\end{equation}

Thus $\phi$ can be interpreted as an electrostatic potential generated by a perfectly conducting object $\mathcal{B}$ that is held at unit potential. (A connection with electrostatics goes back to Berg and Purcell \cite{BP77}; see \cite{book} for review.) The reaction rate $K$ is just the flux 
\begin{equation}
\label{Kdef}
K=D\int_{\partial\mathcal{B}} \nabla n \cdot d\boldsymbol{\sigma}
=-Dn_\infty \int_{\partial\mathcal{B}} \nabla \phi \cdot d\boldsymbol{\sigma}
\end{equation}
Recall that according to electrostatics the total charge on the surface of the equivalent conductor is
\begin{equation}
\label{Q}
  Q=-\frac{1}{4\pi}\int_{\partial\mathcal{B}}  \nabla \phi \cdot d\boldsymbol{\sigma}
\end{equation}
in three dimensions. The total charge on the conductor is related to its capacitance $C$ by $Q=C\phi_{\mathcal{B}}$. In our case $\phi_{\mathcal{B}}=1$ and hence \eqref{Kdef}--\eqref{Q} show that the reaction rate is given by $4\pi n_\infty DC$. Our targets are planar and adsorption is possible only from above, so the proper reaction rate in our case is twice smaller: 
\begin{equation}
\label{K2}
K_2 = 2\pi n_\infty DC
\end{equation}

Generally when the ambient space is $\mathbb{R}^{d+1}$, instead of $4\pi$ we should use $\Omega_{d+1}=2\pi^{(d+1)/2}/\Gamma[(d+1)/2]$, the area of the unit sphere $\mathbb{S}^{d}$ in $\mathbb{R}^{d+1}$. The reaction rate becomes
\begin{equation}
\label{Kd}
K_d = \frac{1}{2}\Omega_{d+1}n_\infty DC
\end{equation}
with factor $\frac{1}{2}$ accounting for the planarity of the target zone. If all dimensions of the target zone are comparable, the capacitance scales as $\ell^{d-1}$ where $\ell$ is the characteristic size of the target zone \cite{LL:EM}. Therefore the reaction rate is proportional to $n_\infty D \ell^{d-1}$ explaining \eqref{eq:ct2} and \eqref{eq:ctd}.

The capacitance of the triangular target zone depends on its shape. For the disk of radius $\ell$, for instance, the capacitance is $C=2\ell/\pi$ \cite{LL:EM}, so $K_2=4n_\infty D \ell$ if the target zone is the disk. Generally for triangular target zones $K_2=A_2 n_\infty D \ell$ with a numerical factor $A_2=O(1)$ depending on the details of the shape of the zone. 

The above analysis is applicable when $d>1$. In the case of the two-dimensional ambient space ($d=1$), equations \eqref{nrt-reaction}, \eqref{nrt-inf}--\eqref{nrt-BC} do not admit a stationary solution. Using the linearity of \eqref{nrt-reaction}, \eqref{nrt-inf}--\eqref{nrt-BC} we still seek solution in the form $n({\bf r}, t)=n_\infty[1-\phi(\mathbf{r},t)]$ and find that $\phi(\mathbf{r},t)$ obeys the diffusion equation
\begin{equation}
\label{DE}
\frac{\partial \phi}{\partial t} = D \nabla^2 \phi 
\end{equation}
and the boundary conditions \eqref{phi:BC}. Fortunately, we do not need to know the full solution as we are interested in the asymptotic long-time behavior. Well inside the depletion zone, $r\ll \sqrt{Dt}$, we can replace the diffusion equation by Laplace equation. Far away from the target zone, $r\gg \ell$, the solution is a combination of two basic solutions of the Laplace equation, a constant solution and $\ln r$. Ensuring the correct match with the inner solution, $\phi\simeq 1$ when $r\sim \ell$, and the outer solution, $\phi\to 0$ when $r> \sqrt{Dt}$, we arrive at
\begin{equation}
\label{phi:log}
\phi(\mathbf{r},t) \simeq 1 - \frac{\ln (r/\ell)}{\ln\sqrt{Dt/\ell^2}}
\end{equation}
To determine the reaction rate \eqref{Kdef} it suffices to compute the flux through the circle of radius $r$ with $\ell\ll r \ll  \sqrt{Dt}$ where \eqref{phi:log} is asymptotically exact. This gives
\begin{equation}
\label{K1}
K_1 = \frac{2\pi n_\infty D t}{\ln(Dt/\ell^2)}
\end{equation}
where we have used the factor $\frac{1}{2}$ accounting for the planarity of the target zone. Thus the density $c(\ell,t)$ of the target zones of length $\ell$ decays almost exponentially,  
\begin{equation}
\label{ct:1}
c(\ell,t) \sim \exp\!\left[-\frac{2\pi n_\infty D t}{\ln(Dt/\ell^2)}\right],
\end{equation}
from which we obtain an interesting decay law 
\begin{equation}
\label{1d:better}
\rho_\text{jam}-\rho(t)\sim  \exp\!\left[-\frac{2\pi n_\infty D t}{\ln(Dt/R^2)}\right]
\end{equation}
in the situation when the substrate is one-dimensional.

\section{Adsorption of Diffusing Squares}
\label{sec:ADSQ}

To analyze the influence of the shape of the diffusing planar objects on the dynamics of the deposition process let us consider the deposition of aligned squares. In the realm of the RSA this problem has been extensively studied, see e.g. \cite{Pal60,Sw81,TTS89,Ziff90,RTTV,Privman91} and reviews \cite{B-Privman,Evans,TT}. 

We assume that the squares are identical, say of the size $R\times R$, and their attachment to the substrate is aligned with $x$ and $y$ axes. We treat the deposition process heuristically using the same approach as in the previous section. The target zones are now rectangles. Long-lived $\ell\times L$ rectangles have sizes smaller than the linear size of the square: $\ell<R$ and $L<R$. Crucially, long-lived rectangles tend to have a very large aspect ratio. The capacitance of an $\ell\times L$ rectangle is
\begin{equation}
\label{CLL}
C\simeq \frac{L}{\ln(2L/\ell)}
\end{equation}
when $L\gg \ell$. This formula gives a correct qualitative behavior even for $L\sim \ell$. More accurate formulas were already known to Maxwell \cite{Maxwell} and the work on this issue is continuing (see e.g. \cite{Jackson} and references therein), but the above is sufficient for heuristic reasoning. 

Using \eqref{K2} and \eqref{CLL} we find
\begin{equation}
c(\ell, L; t) \sim \exp\!\left[- \frac{2\pi n_\infty D L t}{\ln(2L/\ell)}\right]
\end{equation}
The coverage saturates according to
\begin{eqnarray}
\rho_\text{jam}-\rho(t) &\sim& \int_0^R \frac{d\ell}{R}\int_\ell^R \frac{dL}{R}\, c(\ell,L;t)           \nonumber \\
                                   &\sim&    \frac{\ln(n_\infty R D t)}{(n_\infty R D t)^2}           
\end{eqnarray}
This decay law is faster than the $t^{-1}\ln(t)$ decay \cite{Sw81,Privman91} characterizing the asymptotic approach to the jamming coverage in the realm of the RSA of aligned squares.

\section{Adsorption of Diffusing Spheres}
\label{sec:ADS}

Diffusing particles can have various shapes. Diffusing spheres are especially interesting, a suspension of latex spheres being an obvious example, and the behavior of this system has been studied \cite{SJT91,TV,Gray}. In the physically most relevant case of $d=2$, the $t^{-2/3}$ approach to the jammed state has been predicted by Schaaf, Johner and Talbot~\cite{SJT91}. This approach is faster than the $t^{-1/2}$ approach characterizing the RSA of spheres, but slower than the $t^{-1}$ decay law \eqref{eq:ct2} describing the deposition of disks in the diffusion-controlled limit. 

In this section we employ an approach described in Sect.~\ref{sec:ADD} and re-derive the $t^{-2/3}$ behavior. More generally we show that for $d>1$ the jamming exponent is 
\begin{equation}
\label{d>1:ADS}
\sigma = \frac{2}{2d-1}
\end{equation}

In the case of one-dimensional substrate, the exponent $\sigma=2$ predicted by \eqref{d>1:ADS} is correct, but there is also 
an additional logarithmic correction as we argue below:
\begin{equation}
\label{1d:ADS}
\rho_\text{jam}-\rho(t)\sim \left(\frac{\ln t}{t}\right)^2 \quad\text{when}\quad d=1
\end{equation}

At first sight, there seems to be no difference between adsorption of planar disks and spheres. In the RSA framework these two adsorption processes are {\em identical}. Adding diffusion changes the situation. In the case of spheres we can still focus our attention on the deposition of the point particles, the centers of diffusing spheres, into target zones. However to reach the zone, the particle must go through the channel and reach its bottleneck.

\begin{figure}[ht]
\vspace{0cm}
\begin{center}
\includegraphics[width=0.48\textwidth]{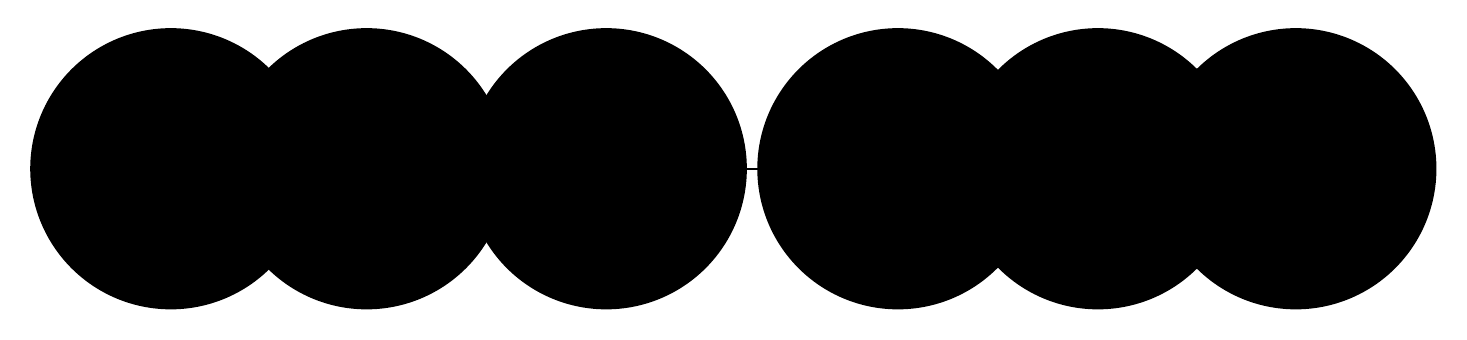}
\caption{Two-dimensional illustration when spheres become (vertical) disks. Each disk represents the excluded region, so its radius twice larger than the radius of diffusing and absorbing particles. The center of the new disk must go through the channel between two central disks and it must reach the bottleneck (the horizontal segment). }
\label{fig-disks_ADS}
  \end{center}
\end{figure}

To establish the large time behavior when the substrate is two-dimensional, $d=2$,  we again replace the diffusion equation by the Laplace equation
\begin{equation}
\label{Laplace}
\nabla^2 n = 0
\end{equation}
Far away from the target 
\begin{equation}
\label{BC:away}
n=n_\infty
\end{equation}
The density must also satisfy the absorbing boundary condition on the target (the bottleneck in Fig.~\ref{fig-disks_ADS}) 
\begin{equation}
\label{BC:target}
n|_\text{on the target}=0
\end{equation}
and the reflecting boundary condition on the boundary of the excluded region. This latter boundary is complicated and random as it depends on the history of the deposition process. When the target zone is very small, however, i.e. $\ell\ll R$, we can significantly simplify the problem. The passage through the very narrow channel is governed by an ``entropic" barrier which provides the dominant contribution to the decay rate. (Problems with entropic barriers are analyzed in several studies, see e.g. \cite{Jacobs,Zwanzig92,entropic06,Bereza07,HS15} and references therein.) Inside such narrow channels the three-dimensional Laplace equation \eqref{Laplace} can be replaced by the quasi-one-dimensional Laplace equation
\begin{equation}
\label{Laplace:Q1d}
\frac{d}{dy}\left[A(y)\frac{dn}{dy}\right] = 0
\end{equation}
where $A(y)$ is the cross-section area of the channel at the height $y$ above the bottleneck. 

In the case of the one-dimensional substrate illustrated in Fig.~\ref{fig-disks_ADS}, if $\ell$ is the width of the bottleneck then the width of the channel is $A(y)=\ell + \frac{y^2}{R}$ when $y\ll R$. In the case of the two-dimensional substrate the cross-section area scales as $A(y)\sim \left(\ell + \frac{y^2}{R}\right)^2$. Using this estimate and integrating \eqref{Laplace:Q1d} we get
\begin{equation}
\label{F:Q1d}
\left(\ell + \frac{y^2}{R}\right)^2\frac{dn}{dy} = F
\end{equation}
Integrating one more time we obtain
\begin{equation}
n=F\sqrt{\frac{R}{\ell^3}}\int_0^{y/\sqrt{R\ell}}\frac{d\eta}{(1+\eta^2)^2}
\end{equation}
Using the boundary condition \eqref{BC:away} we fix the constant 
\begin{equation}
F \sim n_\infty \sqrt{\frac{\ell^3}{R}}
\end{equation}
The flux is given by $DF$. Therefore the density $c(\ell,t)$ of target zones of linear size $\ell$ decays according to 
\begin{equation}
\label{ct:ADS}
\frac{dc(\ell,t)}{dt}\sim - Dn_\infty \sqrt{\frac{\ell^3}{R}} c(\ell,t)
\end{equation}
from which we deduce
\begin{equation}
\label{ADS:2}
\rho_\text{jam}-\rho(t) \sim  (n_\infty RDt)^{-2/3}
\end{equation}

Generally when $d>1$ we use the stationary Eq.~\eqref{Laplace:Q1d} with $A(y)\sim \left(\ell + \frac{y^2}{R}\right)^d$ and arrive at 
\begin{equation*}
\frac{dc(\ell,t)}{dt}\sim - Dn_\infty \sqrt{\frac{\ell^{2d-1}}{R}} c(\ell,t)
\end{equation*}
which leads to to the announced exponent \eqref{d>1:ADS}. More precisely, the coverage saturates according to 
\begin{equation}
\label{ADS:d}
\rho_\text{jam}-\rho(t) \sim  (n_\infty R^{d-1} Dt)^{-\frac{2}{2d-1}}
\end{equation}

The case of the one-dimensional substrate is more subtle. Strictly speaking, one cannot employ the stationary framework. An asymptotically correct results can be established using a simple trick: One formally solves Eq.~\eqref{Laplace:Q1d} to yield 
\begin{equation}
n=F\sqrt{\frac{R}{\ell}}\int_0^{y/\sqrt{R\ell}}\frac{d\eta}{1+\eta^2}
\end{equation}
and then matches this solution to $n_\infty/\ln(R^2 Dt)$ instead of $n_\infty$. This gives 
\begin{equation}
\label{K11}
K_1\sim \sqrt{\frac{\ell}{R}}\,\frac{Dn_\infty}{\ln(R^2 Dt)}
\end{equation}
from which
\begin{equation*}
\ln c(\ell,t)\sim - \frac{n_\infty Dt}{\ln(R^2 Dt)}\, \sqrt{\frac{\ell}{R}} 
\end{equation*}
and
\begin{equation}
\label{ADS:1}
\rho_\text{jam}-\rho(t) \sim  \left[\frac{\ln(R^2 Dt)}{n_\infty Dt}\right]^2
\end{equation}
which is the more accurate version of Eq.~\eqref{1d:ADS}. Instead of the trick used in deriving \eqref{K11} one can establish it repeating the same arguments as in deriving \eqref{K1}. 

A family of one-dimensional RSA-type processes specified by the rates $K_1(\ell)$ of landing into target zones of length $\ell$ has been considered by Tarjus and Viot \cite{TV} who additionally argued that the choice $K_1(\ell)\sim \sqrt{\ell}$ provides an approximation to the diffusion-controlled deposition process. One can verify that choosing such rates leads to the $t^{-2}$ approach to the jamming density  agreeing with \eqref{ADS:1} up to a logarithmic factor. The treatment given in this section accounts for the subtleties of diffusion in two spatial dimensions and therefore differs by an a multiplicative logarithmic factor.

\section{Dense Suspensions}
\label{sec:DS}
 
In this section we show how in principle one can take into account exclusion volume interactions. In the long time limit we have a collection of target zones which are far away from each other and get clogged when a particle touches it. A closely related problem of the survival probability of a trap in diffusive {\em lattice} gases has been recently studied \cite{trap}. Previous work (see \cite{ZKB83,T83,RK84,BO87} and a review \cite{BMS13}) on the survival probability of a trap assumed that diffusing particles do not interact, that is they are random walks (or Brownian particles), but it turned out \cite{trap} that interacting diffusive lattice gases can in principle be treated. Similar behaviors are expected to hold in the present case of {\em continuous} gases of suspended particles. 

Exclusion volume interactions do not change qualitative behaviors. For $d>1$, the jamming exponent remains the same. For $d=1$, the time dependence also remains the same. The new feature is the non-trivial dependence on the volume fraction $n_\infty R^d$ occupied by the particles. 

For instance, in the case of adsorption of planar disks onto the plane 
\begin{equation}
\label{disks-DS}
\rho_\text{jam}-\rho(t) \sim  \frac{\Phi(n_\infty R^3)}{n_\infty RDt}
\end{equation}
and similarly for adsorption of spheres onto the plane 
\begin{equation}
\label{spheres-DS}
\rho_\text{jam}-\rho(t) \sim  \frac{\Psi(n_\infty R^3)}{(n_\infty RDt)^{2/3}}
\end{equation}
When $n_\infty R^3\ll 1$, the exclusion volume interactions can be ignored: $\Phi(\nu)\sim 1$ when $\nu\to 0$, so that \eqref{disks-DS} reduces to \eqref{ct:2d}; similarly $\Psi(\nu)\sim 1$ when $\nu\to 0$ and \eqref{spheres-DS} reduces to \eqref{ADS:d} with $d=2$. 

To compute the functions $\Phi(\nu)$ and $\Psi(\nu)$ in the case of dense suspensions seems impossible, although the formal scheme of the computation exists \cite{trap} and it is based on a macroscopic fluctuation theory (see \cite{MFT:rev} for a review). In the long time limit it suffices to consider a single target zone. The logarithm of the probability that it remains uncovered at time $t$ is asymptotically
\begin{equation}
\label{S:MFT}
-\ln S \simeq \frac{1}{2}\,t\int d{\bf r}\,\frac{D^2(n)}{\sigma(n)}\,(\nabla n)^2
\end{equation}
when $d>1$. The integral in Eq.~\eqref{S:MFT} is taken over the $(d+1)-$dimensional ambient space outside the target zone. The integrand in \eqref{S:MFT} contains two transport coefficients, the diffusion coefficient $D(n)$ and the mobility $\sigma(n)$. The density  $n({\bf r})$ is determined by the solution of a {\em stationary} partial differential equation
\begin{equation}
\label{n:PDE}
\nabla^2 n + \left(\frac{D'}{D}-\frac{\sigma'}{2\sigma}\right) (\nabla n)^2 = 0
\end{equation}
where $D'=dD/dn$ and $\sigma'=d\sigma/dn$. The boundary conditions are \eqref{BC:away}--\eqref{BC:target} and the the reflecting boundary condition on the boundary of the excluded region. The case of planar disks is again particularly simple as we do not need the latter boundary condition, we can instead analyze \eqref{n:PDE} in $\mathbb{R}^{d+1}$. 

It is usually impossible to solve the non-linear partial differential equation \eqref{n:PDE} subject to the boundary conditions \eqref{BC:away}--\eqref{BC:target}. For a few simple lattice gas models explicit analytical solutions were found in \cite{trap}, see also  \cite{in-trap-1, in-trap-2}, but one should keep in mind that for almost all interacting gases the transport coefficients $D(n)$ and $\sigma(n)$ are unknown. Even if we one finds a simple interacting gas with density-independent diffusion coefficient, the mobility is still very hard to determine---one needs to know the free energy, and in two and higher dimensions it is essentially impossible for any gas with exclusion interactions.

\section{Discussion}
\label{sec:Discussion}

We analyzed the irreversible deposition of diffusing particles onto the substrate. The substrate eventually gets clogged, but even the simplest characteristic of the resulting jammed state, the filling fraction, has not been computed even in the simplest settings. A popular framework known as the random sequential adsorption (RSA) is convenient for numerical implementation and it has led to interesting predictions about the large time asymptotic behaviors. The RSA is essentially a computer algorithm---the motion of particles and interactions between them are totally ignored, deposition events are consecutively attempted and an event is successful if an arriving particle does not overlap with already deposited ones, otherwise the deposition event is discarded. 

\begin{table}[t]
\begin{tabular}{|c|c|c|c|c|c|}
\hline 
 Process    & \multicolumn{2}{|c|}{RSA}   & \multicolumn{3}{|c|}{Diffusion-controlled}   \\
\hline
substrate    & disks or       & aligned         & disks     & aligned                & spheres \\
dimension  & spheres      &  squares       &              & squares               &  \\
\hline
$d=2$        & $t^{-1/2}$   & $t^{-1}\ln t$    & $t^{-1}$        & $t^{-2}\ln t$    & $t^{-2/3}$\\
\hline
$d=1$        & $t^{-1}$      & $t^{-1}$         & $e^{-t/\ln t}$   & $e^{-t/\ln t}$   & $t^{-2}(\ln t)^2$  \\
\hline
\end{tabular}
\caption{The asymptotic approach to the jamming coverage for the RSA and for the diffusion-controlled deposition process. In one dimension, the aligned squares and disks are just segments, so the behavior is the same.}
\end{table}

In deposition processes that occur in Nature, particles usually diffuse in the solution above the substrate. One may guess that diffusion accelerates the approach to the jamming state. This is correct and Table I compares basic predictions for diffusion-controlled deposition processes with corresponding kinetic laws for the RSA. 

Diffusion-controlled deposition processes are genuinely infinite-particle systems, while by construction the RSA is effectively a single-particle system, albeit with memory. The RSA case has led to the guess that the large time behaviors may be tractable. We additionally utilized the chief idea employed in Refs.~\cite{P80,Sw81} in the case of the RSA, namely we wrote a rate equation describing the decay of the density of the target zones. To derive a reaction rate appearing in this equation we also used classical ideas essentially going back to Smoluchowski \cite{SM16} who developed them in the context of coagulation. 

We employed heuristic arguments, but predictions appear to be asymptotically exact. Proving these predictions is a challenge, very few such proofs have been constructed even for much more simple diffusion-controlled infinite-particle systems. Diffusion-controlled deposition processes are more difficult to simulate than the RSA, and even for the RSA extracting the long time kinetic behavior is challenging since the jamming coverage is not known and finite-size corrections are not fully understood \cite{ZT13,CZ18}. Still, simulations of diffusion-controlled deposition processes have been done \cite{Gray} and they seem to confirm the $t^{-2/3}$ behavior arising in the case of diffusing spheres. In the presumably simpler case of the one-dimensional substrate, the asymptotic behaviors (see Table I) involve logarithms which probably makes numerical confirmation more challenging. 

We  investigated the deposition of disks, spheres and aligned squares.  It would be interesting to consider objects of different shapes. The shape and the orientational freedom of the depositing objects may affect the asymptotic behaviors. For instance, one would like to explore the diffusion-controlled deposition of planar ellipses (planar squares) in the case of the isotropic deposition, i.e., assuming that the ellipses (squares) undergo both translational and rotational diffusion. In the realm of the RSA the asymptotic behaviors depend on symmetries of the objects and the orientational freedom \cite{TT,TTS89,Ziff90,RTTV,Privman91}, but in the isotropic case the universal $t^{-1/3}$ behaviors emerges for squares (and generally for rectangles) and for ellipses with arbitrary non-zero eccentricity \cite{TTS89,Ziff90,RTTV}.

The deposition process never ends if the depositing objects have no width. For the RSA of needles, for instance, the density of needles per unit area exhibits a remarkable growth law $t^{\sqrt{2}-1}$ characterized by an irrational exponent \cite{TV91}, see also \cite{KB94,JTT94} for further results and analyses of related fragmentation problems. The diffusion-controlled deposition of needles is an intriguing open problem. 

In Sects.~\ref{sec:ADD}--\ref{sec:ADS} we ignored interactions between diffusing particles.  At first sight the inclusion of the simplest exclusion volume interactions already makes the system intractable, e.g. the equilibrium statistical mechanics of such infinite systems is unknown. However, the hydrodynamic behavior is in principle simple: it is governed by a diffusion equation with density-dependent diffusion coefficient $D(n)$.  To analyze diffusion-controlled deposition processes in addition to $D(n)$ one must know another transport coefficient, the mobility $\sigma(n)$. In Sect.~\ref{sec:DS} we showed that in the most interesting case when $d>1$ the inclusion of interactions forces one to seek a solution of the non-linear partial differential equation \eqref{n:PDE} instead of the Laplace equation. Thus one can proceed, at least numerically, if one knows $D(n)$ and $\sigma(n)$. For a few simple {\em lattice} gases, including e.g. the symmetric exclusion process, the transport coefficients are known and an unexpected ansatz  \cite{trap} allows one to treat the non-linear partial differential equation \eqref{n:PDE}.

\bigskip\noindent
I am grateful to Julian Talbot for interesting discussions.

\end{document}